\documentclass[aps,prl,twocolumn,superscriptaddress]{revtex4-2}
\usepackage[english]{babel}
\usepackage{amsmath,amssymb,braket,pifont,amsthm,lipsum}
\usepackage{amsfonts}
\usepackage[pdfborder={0 0 0}, colorlinks=true, urlcolor=blue,linkcolor=blue, citecolor=blue]{hyperref}
\usepackage{color}
\usepackage[dvipsnames]{xcolor}
\usepackage{graphicx}
\usepackage{epstopdf}
\usepackage{float}
\usepackage{subfigure}
\usepackage{ulem}
\usepackage{url}
\usepackage{lineno}
\makeatletter

\definecolor{color1}{rgb}{0.0,0,1.0}

\definecolor{color2}{rgb}{0.7,0,0.2}

\usepackage{babel}
\makeatother

\begin{document}

\title{Dissipative Preparation of Correlated Quantum States in Dipolar Rydberg Arrays}

\author{Mingsheng Tian}
\affiliation{Department of Physics, The Pennsylvania State University, University Park, Pennsylvania, 16802, USA}
\author{Zhen Bi}
\affiliation{Department of Physics, The Pennsylvania State University, University Park, Pennsylvania, 16802, USA}
\affiliation{Center for Theory of Emergent Quantum Matter, The Pennsylvania State University, University Park, Pennsylvania 16802, USA}
\author{Thomas Iadecola}
\affiliation{Department of Physics, The Pennsylvania State University, University Park, Pennsylvania, 16802, USA}
\affiliation{Center for Theory of Emergent Quantum Matter, The Pennsylvania State University, University Park, Pennsylvania 16802, USA}
\affiliation{Institute for Computational and Data Sciences, The Pennsylvania State University, University Park, Pennsylvania 16802, USA}
\affiliation{Materials Research Institute, The Pennsylvania State University, University Park, Pennsylvania 16802, USA}
\author{Bryce Gadway}
\email{bgadway@psu.edu}
\affiliation{Department of Physics, The Pennsylvania State University, University Park, Pennsylvania, 16802, USA}

\begin{abstract}
Preparing correlated quantum states is essential for emerging technologies, but remains challenging in many-body systems. Here we propose a dissipative protocol that engineers nonreciprocal, energy-selective transitions to steer dipolar quantum systems toward desired many-body states. This is realized by introducing two types of controllable dissipative auxiliary atoms that act as nonreciprocal excitation and de-excitation channels, respectively, enabling a directional walk in Hilbert space. This approach enables stabilization of states across the many-body spectrum, not limited to the ground state and requiring no \textit{a priori} knowledge of the Hamiltonian. Our approach is designed for neutral atoms in dipolar Rydberg arrays, but applies broadly to setups with similar capabilities, providing a flexible and scalable framework for state preparation in programmable platforms.
\end{abstract}

\maketitle
Preparing and controlling quantum states is a central objective in quantum science, 
particularly in many-body quantum physics.
The emergence of strong correlations and entanglement, manifested both in equilibrium 
ground states and out-of-equilibrium dynamics, provides a key resource for quantum technologies.
Quantum simulators offer an attractive route to explore quantum state engineering due to their precisely controlled and tunable interactions and dynamics~\cite{Georgescu2014,altman2021}. 
In current experiments, low-entropy correlated states are typically reached 
adiabatically by slowly tuning the system Hamiltonian through a quantum phase transition~\cite{greiner2002-adia,friedenauer2008,simon2011-adia,mazurenko2017-adia,monroe2021-adia,wang2024-adia,leonard2023-adia,impertro2025-adia,deleseleuc2019-adia,ebadi2021-adia,semeghini2021-lukin}. 
However, this approach is challenged in many scenarios: it often requires the addition of tailored symmetry breaking to reach many ground states~\cite{chen2023-adia,cohen2014,wu2025}, is hindered by gap closings across continuous phase transitions in finite-time preparation~\cite{Sachdev_2011,keesling2019a,zeng2025}, and is challenged by competition from heating in Floquet-engineered systems~\cite{dalessio2014,goldman2014b,weitenberg2021a,yang2022}.

An alternative approach, known as dissipative state preparation~\cite{kraus2008a,verstraete2009,harrington2022,fazio2025-open,lin2025a-open,zhan2026-open,dorstel2025,feldmeier2026a}, provides a direct 
route for state preparation without crossing a critical point. 
This approach evolves the system
under engineered dissipation and encodes the target state as the stationary state solution of a governing master equation~\cite{masterequation}. 
Such schemes rely on designed dissipative channels and  are  well suited to highly controllable synthetic quantum platforms~\cite{diehl2008,barreiro2011,mi2024,li2024,ma2019}. For example, recent theoretical and experimental progress in superconducting circuits has enabled the dissipative preparation of small-scale microwave-photon Mott insulators~\cite{ma2017,ma2019}. Extending such approaches to other platforms may enable the dissipative preparation of scalable and complex many-body states.

Rydberg atom arrays offer a promising platform for realizing, tuning, and detecting many-body states due to their flexible geometries and strong, controllable interactions~\cite{adams2019-rydbergreview,browaeys2020-adia}. 
Advances in the precise trapping and manipulation of neutral atoms in programmable arrays have enabled the exploration of correlated quantum phases~\cite{deleseleuc2019-adia,ebadi2021-adia,chen2023-adia,semeghini2021-lukin,tian2025} and 
nonequilibrium dynamics~\cite{bernien2017-adia,bluvstein2021b-dynamics,tom1014-dynamics,chen2025d}. 
More recently,  progress in engineering controlled loss channels provides a 
natural setting for studying open quantum many-body physics~\cite{chen2025c,zhang2025b,begoc2025,wang2026}. 
It is therefore timely to ask whether, e.g., the pioneering approach to dissipation engineering in superconducting qubit arrays~\cite{ma2017,ma2019,li2024} can be scaled and improved in large arrays of intrinsically identical atoms, ions, or molecules.

\begin{figure}[tb]
    \centering
    \includegraphics[width=0.95\linewidth]{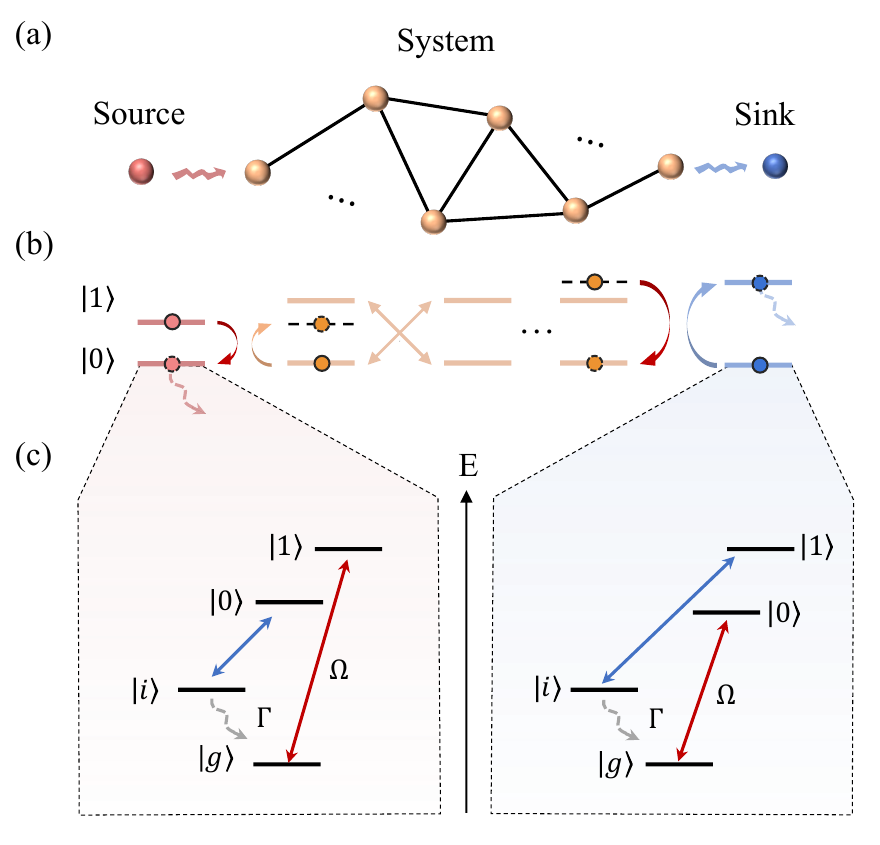}
    \caption{\textbf{Engineered dissipative stabilization scheme.}
(a)~An interacting many-body system coupled to two types of engineered auxiliary atoms, which act as ``source'' and ``sink'' to excite (de-excite) particles to (from) the ``system.''
(b)~Physical realization by dipolar spin-exchange processes in Rydberg arrays. 
Auxiliary atoms are engineered with state-dependent dissipation: the source (sink) has strong decay from state $\ket{0}$ ($\ket{1}$), leading to asymmetric rates of spin exchange with the system. 
The source (sink) atoms' state energies are also shifted with respect to the natural $\ket{0} \leftrightarrow \ket{1}$ transition, controlling the energy-dependent flow of excitations into (out of) the system.
(c)~Optical control of auxiliary atoms. A selected Rydberg state is resonantly coupled to a short-lived intermediate state $\ket{i}$ with Rabi frequency $\Omega_0$ (blue), acquiring an effective decay at rate $\gamma\sim \Omega_0^2/\Gamma$ (if 
$\Gamma \gg \Omega_0$).
Continuous optical driving (red) between ground and Rydberg states provides the persistent and tunable pumping for auxiliary atoms.}
    \label{fig1}
\end{figure}

Here, we present dissipative protocols to stabilize 
many-body correlated states in Rydberg arrays. 
Specifically, we consider two types of auxiliary atoms (referred to as ``source'' and ``sink'') coupled to a system, featuring state-selective driven-dissipative channels and tunable transition frequencies, which enable nonreciprocal particle excitation and de-excitation at chosen transition energies [see Fig.~\ref{fig1}(a)].
By controlling the
``source" and ``sink" spectra, generic initial states are directed to target many-body states through engineered dissipation channels.
As concrete applications, we provide a framework for preparing many-body ground states with different particle fillings in generic interacting systems, without requiring a \textit{priori} knowledge of the many-body spectrum, and numerically verify their efficiency in the interacting dipolar XY model. 
Beyond ground states, we further provide spectral engineering protocols for preparing excited many-body states, enabling the exploration of excited state many-body physics. 
We thus establish a scalable framework for energy-resolved dissipative many-body state preparation in dipolar spin arrays.

\textit{Our protocol.}---We consider a Rydberg atom array, 
where each atom hosts four relevant states: the ground state $\ket{g}$, a low-lying excited state $\ket{i}$ that decays to $\ket{g}$, and two selected Rydberg states $\ket{0}$ and $\ket{1}$, which encode an effective spin-$1/2$ degree of freedom and have negligible intrinsic dissipation. 
Dipolar interactions between Rydberg states are used to engineer the Hamiltonian, for example a spin-$1/2$ XY model,
$
\hat{H}_{\rm S}=-\sum_{i j}(V_{ij}\,\hat{S}_i^+ \hat{S}_j^- + \mathrm{h.c.)},
$
where $\hat{S}_i^+$ denotes the spin-raising operator and $V_{ij}\propto 1/|r_i-r_j|^3$) is the interaction strength between atoms $i$ and $j$.
Without restricting to a specific model, we express the system Hamiltonian in its eigenbasis as
$
\hat{H}_{\text{S}} = \sum_{\lambda} \lambda\, |\lambda\rangle\langle \lambda| .
$
The raising operator in this basis is
$
\hat{S}_i^{+} = \sum_{\lambda,\lambda'} s_{i,\lambda\lambda'} |\lambda\rangle\langle\lambda'|
$,
with $s_{i,\lambda\lambda'}=\langle \lambda|\hat{S}_i^+|\lambda'\rangle$.

To control such a system, we couple it to auxiliary atoms via a flip–flop dipolar interaction, 
\begin{equation}\label{eq:hsa}
        \hat{H}_{\rm SB}=\sum_{ij}J_{ij}e^{-i\Delta_j t} \hat{\sigma}_j^- \hat{S}_i^++\rm{h.c.} \ .
\end{equation}
Here, $\sigma_j^-$ is the spin-lowering operator for auxiliary atom $j$ and $\Delta_j$ is the 
detuning of its resonance [$\ket{0} \leftrightarrow \ket{1}$ in Fig.~\ref{fig1}(b) and (c)], relative to that of the system atoms.
This detuning controls the resonance condition and introduces the phase factor $e^{-i\Delta_j t}$ for spin exchange.

We recast Eq.~(\ref{eq:hsa}) in a rotating frame via $\hat U=e^{-i \hat{H}_{\rm S}t}$,
\begin{equation} \label{eq2}
\hat{\widetilde{H}}_{\rm SB}
= \sum_{ij} \sum_\omega J_{ij} e^{i(\omega-\Delta_j)t}\,\
\hat{\sigma}_j^- \hat{S}_i^+(\omega) + \text{h.c.},
\end{equation}
with
$
\hat{S}_i^+(\omega)
=\sum_{\lambda-\lambda'=\omega} s_{i,\lambda\lambda'}
\ket{\lambda} \bra{\lambda'} 
$
denoting the raising operator at a specific frequency. 
Under the rotating-wave approximation, 
oscillatory off-resonant terms average out, yielding an effective transition rate $R_{\lambda\leftrightarrow\lambda'}\propto \delta(\omega-\Delta_j)$.
Thus,  
control of $\Delta_j$ allows for selective addressing of specific Bohr-frequencies of the system.

Central to our stabilization protocol is the realization of nonreciprocal coupling terms, i.e.,
$R_{\lambda\to\lambda'}\neq R_{\lambda'\to\lambda}$, which enables directional preparation of the target state through nonreciprocal, energy-selective state transitions.
To this aim,
we introduce two types of state-selective dissipation for the auxiliary atoms. One class of ``source'' atoms are engineered to have rapid decay from $\ket{0}$, while another class of ``sink'' atoms have large decay from $\ket{1}$  (Fig.~\ref{fig1}(b)). This is realized by resonantly coupling selected Rydberg states to a short-lived intermediate state $\ket{i}$, as illustrated in blue lines of Fig.~\ref{fig1}(c). Owing to rapid spontaneous emission from $\ket{i}$, this coupling induces a tunable effective decay rate ($\sim \Omega_0^2/\Gamma$~\cite{chen2025c}) from the Rydberg state back to the ground state.

In the presence of such dissipation, the spin-exchange processes in Eq.~(\ref{eq2}) become effectively nonreciprocal.
The source atoms with fast decay from $\ket{0}$ are more likely to undergo exchange with the system atoms by giving up an excitation, with the reverse process being strongly suppressed.
That is, once the source atom reaches the state $\ket{0}$ it rapidly decays and cannot efficiently absorb an excitation from the system.
As shown in Fig.~\ref{fig1}(b), such a nonreciprocal flip-flop interaction can therefore excite a single particle with a specific frequency in the system. 
Crucially, optical driving of the source atoms ($\Omega\ket{g}\bra{1}+\mathrm{h.c.}$) replenishes the population of $\ket{1}$, allowing this process to repeat and allowing such auxiliary atoms to act as persistent ``sources'' of excitations~\cite{ma2017,ma2019}. Conversely, the auxiliary atoms experiencing induced decay from $\ket{1}$ [right panel in Fig.~\ref{fig1}(c)] can be dissipatively prepared with high probability in the state $\ket{0}$ (assuming an analogous optical driving of these atoms, $\Omega \ket{g}\bra{0}$ + h.c.), allowing them to serve as ``sinks'' that can efficiently absorb excitations from the system atoms.

\begin{figure*}[tb]
    \centering
    \includegraphics[width=1\linewidth]{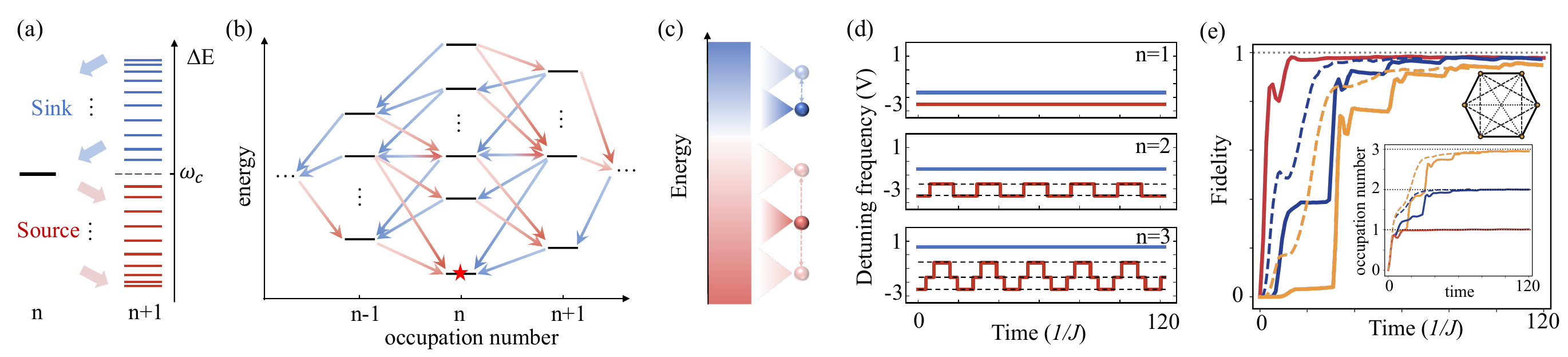}
    \caption{
    \textbf{Dissipative preparation of $n$-filled ground states.}
    (a)~Energy-selective pumping and depumping processes. 
    Transitions from $n$ to $n+1$ with frequency below $\omega_c$ are predominantly driven by the source atoms, whereas transitions above $\omega_c$ are dominated by the sink atoms. 
    $n=\sum_i \hat{S}_i^\dagger \hat{S}_i$ denotes the total occupation number.
    (b)~Schematic illustration of the preparation protocol as a directed walk in the many-body Hilbert space. 
    Black lines denote the energy levels in the rotating frame defined by $\omega_c$. 
    Red and blue arrows represent excitation and de-excitation processes, respectively.
(c)~Detuning protocols for the source (red) and sink (blue) atoms used to stabilize ground states, with coverage of the desired spectral range accomplished either through multiple auxiliaries or through slow temporal modulation.
(d,e)~Example of preparing ground states with different fillings in a dipolar XY model. 
(d)~Spectral engineering of the source (red) and sink (blue) atoms for different-fillings ground state preparation. 
A broad spectral range is achieved either by slowly rastering a single auxiliary detuning in time (solid lines) or by using multiple auxiliaries with different detunings (dashed lines).
(e)~Time evolution of the fidelity between the prepared state and the target ground state with $n=1$ (red), $2$ (blue), and $3$ (yellow). 
The inset shows the corresponding time evolution of the total excitation number starting from the vacuum state.
The yellow and blue dashed lines show the corresponding results by using multiple auxiliaries, indicating faster preparation.
Parameters: $\gamma/J=0.65$, $\Omega/J=0.12$, $V/J=20$, $N=6$ with hexagon structure.
Here, $V$ denotes the nearest neighbor dipolar interaction and $J$ is the system-auxiliary coupling.
The dynamics are simulated with the quantum trajectory method based on 5000 trajectories.
}
    \label{fig2}
\end{figure*}

With the final ingredient of optical AC Stark shifting~\cite{deleseleuc2017}, so as to shift the $\ket{0} \leftrightarrow \ket{1}$ resonance frequency condition of auxiliary atoms in a locally controlled way, the source and sink atoms can act as energy-selective driven and dissipative channels, respectively.
By control of the detuning values ($\Delta_j$), one can selectively inject or remove excitations at specific Bohr frequencies of the system, enabling energy-resolved nonreciprocal transitions in the system Hilbert space.

\textit{Stabilizing ground states}.---Building on this, the main idea of our state preparation protocol is 
to separately engineer the spectral response of the auxiliary ``source'' and ``sink'' atoms through their detunings $\Delta_j$.
In the prototypical case of preparing ground states, the source detunings
are arranged to cover lower transition energies, while the sink atoms have coverage at higher energies, as sketched in Fig.~\ref{fig2}(a). As a result, there exists a critical energy $\omega_c$ separating two regimes. For transition energies below $\omega_c$, the system is more likely to absorb ``source'' excitations, being directed from eigenstates with occupancy $n$ to $n+1$ ($\Delta n = +1$). Conversely, above $\omega_c$, the ``sink'' coupling dominates and drives system transitions with $\Delta n = -1$.

To characterize its stabilization mechanism, we work in the rotating frame defined by
$
\hat U=e^{i\omega_c t\sum_i \hat n_i},
$
where an eigenstate in the $n$-particle sector with energy $\lambda_n$ is shifted to
$
E_n^{\rm rot}=\lambda_n-n\omega_c$.
This transformation effectively introduces a tunable energy offset proportional to the particle number, so that different $n$-fillings are relatively tilted by $\omega_c$.
For a transition $n\to n+1$,
when $\lambda_{n+1}-\lambda_n<\omega_c$, the transition (red arrows in Fig.~\ref{fig2}(b)) is dominated by the source and favors adding a particle to the system and lowering the rotating-frame energy. In contrast, when $\lambda_{n+1}-\lambda_n>\omega_c$, the transition (blue arrows) is dominated by the sink and favors removing a particle from the system and also lowering the rotating-frame energy.
As a result, the dynamics becomes a directed flow in Hilbert space, as schematically illustrated in Fig.~\ref{fig2}(b), with all trajectories biased toward the states selected by minimizing the energy $E_{n}^{\rm rot}$. The corresponding ground-state filling is determined by
\begin{equation}
n_{*}=\arg\min_n \left(\lambda_n-n\omega_c\right).
\label{eq_minenergy}
\end{equation}
By tuning $\omega_c$, one can continuously shift the relative energies of different particle-number sectors. Decreasing $\omega_c$ favors smaller $n$, whereas increasing $\omega_c$ favors larger $n$. In this way, the minimum in Eq.~(\ref{eq_minenergy}) can be tuned to different fillings, enabling selective stabilization of ground states at different fillings.

To demonstrate it, we consider a dipolar interacting XY model and explore such control in Fig.~\ref{fig2}(d,e).
Specifically,
we simulate the dynamics of a small hexagon XY model with long-ranged dipolar couplings from an initial vacuum state and
show that
variation of $\omega_c$ across the many-body spectrum allows preparation of ground states with tunable filling.
One approach to setting $\omega_c$ through the (time-averaged) auxiliary spectra is to use 
a single source and sink
and implement a slow temporal scan of $\Delta$. Such a detuning protocol is illustrated in Fig.~\ref{fig2}(d), and Fig.~\ref{fig2}(e) shows the resulting time-evolving fidelity between the prepared state and ground states at different target fillings. 
Alternatively, a fleet of atoms with static, dispersed $\Delta$ values (dashed lines in Fig.~\ref{fig2}(d)) can effectively define the auxiliary spectra.
The resulting fidelity dynamics, dashed lines in Fig.~\ref{fig2}(e), reveal that this approach improves the rate of dissipative preparation, avoiding the ``dead time'' associated with time-dependent scanning. While experiments will thus ideally use a number of auxiliaries that scales with the system size, our numerics are limited to small numbers.
Although we have focused on an XY model, we emphasize that the described protocol is entirely general. 
For example,
we show in the Supplement~\cite{supp} how to stabilize ground states with different fillings in the strongly interacting hard-core Bose--Hubbard model with magnetic flux, where the ground state corresponds to fractional quantum Hall-like states.

\textit{Stabilizing non-ground states}.---In many settings, the ability to prepare excited states is as important as preparing the ground 
state of a Hamiltonian. Excited states govern dynamical and spectroscopic properties of 
quantum systems and are central to nonequilibrium phenomena~\cite{giannetti2016,bravyi2006,calabrese2006,cheneau2012,chen2025d}.  
In certain model systems, excited many-body states may play host to phenomenology~\cite{Huse13} and phase transitions~\cite{Iadecola18,Schecter18,cejnar2021} distinct from those of the ground state.
However, controlled preparation of excited 
states poses a significantly greater challenge than ground-state preparation~\cite{higgott2019,Zhang21}. For bosonic systems in particular (including spin systems), excited states hosting intriguing properties may be unstable to decay~\cite{bracamontes2022a,wang2021,sedrakyan2015b}.

\begin{figure}[tb]
    \centering
    \includegraphics[width=1\linewidth]{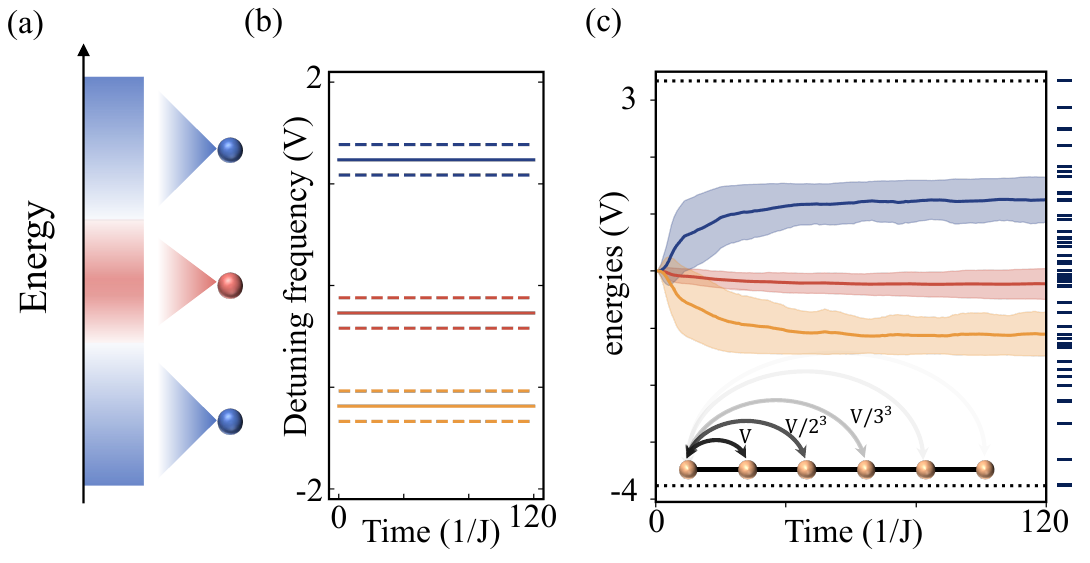}
    \caption{
\textbf{Dissipative preparation of non-ground steady states.}
(a)~Sketch of energy-selective stabilization of intermediate excited states.
(b)~Detuning protocols applied to a single source (solid) and two sink (dashed) auxiliaries. From top to bottom are three source detuning protocols for the preparation of different energy states.
(c)~Time evolution of the prepared-state energy. The blue, red, and yellow curves correspond to the top-to-bottom detuning protocols in (b), respectively. Shaded bands indicate the energy standard deviation. The narrow strip on the right shows the corresponding many-body spectrum of a six-site dipolar XY chain.
Other parameters are the same as in Fig.~\ref{fig2}(e).
}
    \label{fig3}
\end{figure}

Our protocol allows stabilizing the highest-energy state in a way analogous to ground-state preparation by simply exchanging the detuning configurations of the source and sink auxiliaries. More generally, to stabilize intermediate excited states, we can engineer the auxiliary spectra so that source atoms couple to the system more strongly over an intermediate energy window $[\omega_-,\omega_+]$, while sink atoms couple more strongly at both lower and higher energies, as illustrated in Fig.~\ref{fig3}(a).  This realizes a bidirectional energy filtering mechanism that directs the dynamics toward the selected window. By tuning the bounds $\omega_-$ and $\omega_+$, different intermediate states can be prepared.

For quadratic fermionic-like models, the Hamiltonian can be diagonalized in terms of independent quasiparticles (e.g., via a Bogoliubov transformation). In this basis, quasiparticles with energies in $[\omega_-,\omega_+]$ are preferentially excited by the source atoms, while quasiparticles outside this window are de-excited by the sinks.
The steady states therefore correspond to the specific many-body eigenstates whose quasienergies lie in the selected energy window.
For general interacting systems, the characteristic of steady state cannot be analytically captured via quasiparticles excitation and de-excitation.
To characterize it, we consider a 1D dipolar XY model as our system Hamiltonian $\hat{H}_{\rm S}$ and numerically calculate the system energy evolution on time.
As shown in Fig.~\ref{fig3}(b,c), by tuning the source and sink spectra, the state can be prepared into distinct non-ground states with different energies. 
This may enable investigations of energy-dependent structure across the many-body spectrum.
For example, in disordered systems like the random field Heisenberg chain~\cite{znidaric2008,luitz2015b} or quantum Ising chain~\cite{kjall2014}, one can probe the energy-resolved distinction between thermal eigenstates consistent with the eigenstate thermalization hypothesis (ETH)~\cite{Deutsch91,Srednicki94,DAlessio2016} and nonthermal eigenstates associated with many-body localization~\cite{schreiber2015,abanin2019,alet2018},
the coexistence of which
has been discussed in the context of many-body mobility edges~\cite{znidaric2008,pal2010,kjall2014,laumann2014,luitz2015b,deroeck2016}.
It may also provide a route to estimate microcanonical statistical averages in ETH-obeying systems, where superpositions and mixtures of eigenstates in a sufficiently narrow energy window can act as experimentally accessible surrogates for complex many-body wavefunctions~\cite{Pollock23}.

\textit{Experimental feasibility.—}While Rydberg state-specific loss is not natural, it can be engineered by optically coupling Rydberg states to short-lived intermediate states $\ket{i}$ (depicted in Fig.~\ref{fig1}(c)). This induces a tunable state-selective decay rate $\gamma\sim\Omega^2 / \Gamma$~\cite{chen2025c} when the decay rate $\Gamma$ far exceeds the coupling rate $\Omega$. The induced decay can return the atom to a single ground state $\ket{g}$ by judicious choice of state $\ket{i}$, for example the stretched state $\ket{nL_J,F,m_F} = \ket{5P_{3/2},3,3}$ of $^{87}$Rb or the $^1P_1$ state of $^{174}$Yb. Finally, while selection rules limit the optical coupling of the state $\ket{i}$ to Rydberg levels of opposite parity, the Rydberg states $\ket{0}$ and $\ket{1}$ of the auxiliary atoms can both be effectively coupled by setting one to be a bare Rydberg level (e.g., $\ket{0} \equiv \ket{62s}$) and the other to be an inert (i.e., without Ising-like self-interactions) dressed superposition state of opposite-parity Rydberg levels (e.g., $\ket{1} \equiv (\ket{62p} + \ket{64s})/\sqrt{2}$) strongly hybridized by resonant microwave dressing. We remark that the three types of atoms---system, source, and sink---can all be of the same atomic species, with their roles defined solely through local optical addressing. Alternatively, multiple species can be utilized under F\"orster resonance conditions~\cite{Forst-Saff,Anand2024}, reducing the requirements for local addressing.
 
The detuning of the auxiliary ``sink" and ``source" atoms can be implemented by applying state-selective light shifts with addressing beams~\cite{deleseleuc2017}. The detuning $\Delta_i$ for sink or source can be tuned over a broad range---from a few kHz to even tens of MHz---by varying the intensity and the detuning of the addressing lasers.
By modulating the Rabi frequency $\Omega_{\rm addr}$ of the addressing beams via the laser intensity using fast acousto-optic or electro-optic (amplitude) modulators, one can generate the desired time-varying $\Delta$ profiles. 
As this scheme is based on optical fields, the control can be implemented with high spatial resolution, thereby realizing atom-resolved time-dependent detuning on sink and source. 
Multiple auxiliaries can be arranged, either embedded within a two-dimensional array or by utilizing layered geometries~\cite{jwp-array}, e.g. with a plane of system atoms surrounded by source and sink atoms. Control over system–auxiliary coupling strength can be provided by spatial separation and the distance dependence of dipolar interactions (i.e., in multi-layer or multi-strip scenarios), or via state selection.

Beyond Rydberg arrays, this strategy can likely be adapted to other dipolar systems, 
such as polar molecule arrays~\cite{yan2013a,Cheuk-Array,Bao-array,Ruttley2025} and dipolar spin arrays in solids~\cite{rovny2018,davis2023}, which may provide larger temporal windows for dissipative stabilization as compared to Rydberg atoms.
As it takes inspiration from work on transmon qubits~\cite{ma2017,ma2019}, this protocol is also not restricted to dipolar systems, but can be implemented in generic systems that support excitation exchange. For example, in trapped ions, where spin exchange is mediated through virtual excitations of a phonon bus~\cite{richerme2014a,jurcevic2014a,blatt2012}, there may be added opportunities for stable dissipative preparation~\cite{Lin2013} based on independent motional cooling by auxiliary ions~\cite{symp-ions}.

\textit{Conclusion.—}
We have introduced a many-body state stabilization approach for dipolar spin systems. 
By exploiting controllable dissipation and optically tunable light shifts of the auxiliary atoms, we engineer nonreciprocal, energy-selective excitation and de-excitation channels in the many-body spectrum. 
Building on this, we construct a framework to stabilize many-body ground states with tunable fillings in a generic Hamiltonian without \textit{a priori} knowledge of the many-body spectrum. We also give a protocol to stabilize the excited states at a finite energy window, offering a route to exploration of many-body excited physics, such as the existence of thermal and localized many-body states in high-excited states~\cite{znidaric2008,pal2010,kjall2014,laumann2014,luitz2015b,deroeck2016}.
Our work therefore provides a scalable and controllable framework for preparing correlated states of many-body systems, with added benefits such as stabilization against Floquet heating.

\acknowledgments{
We thank Jon Simon for stimulating discussions. This work (M.~T. and B.~G.) was supported in part by the AFOSR MURI program under Agreement No.~FA9550-22-1-0339 and in part by the Gordon and Betty Moore Foundation (Grant No.~13778). M.~T. was partially supported by a Rising Researcher Award from Penn State’s Institute for Computational and Data Sciences (RRID:SCR\_025154). We also acknowledge support from the National Science Foundation under Grants No.~DMR-2339319 (Z.B.) and No.~DMR-2611305 (T.I.). Z.B.~and T.I.~acknowledge partial support from a Quantum SuperSEED grant (ICDS\_QS25\_029093) from the Institute for Computational and Data Sciences at the Pennsylvania State University. 
}


%

\end{document}


\title{Supplemental Material: Dissipative Preparation of Correlated Quantum States in Dipolar Rydberg Arrays}    

\author{Mingsheng Tian}
\affiliation{Department of Physics, The Pennsylvania State University, University Park, Pennsylvania, 16802, USA}
\author{Zhen Bi}
\affiliation{Department of Physics, The Pennsylvania State University, University Park, Pennsylvania, 16802, USA}
\affiliation{Center for Theory of Emergent Quantum Matter, The Pennsylvania State University, University Park, Pennsylvania 16802, USA}\author{Thomas Iadecola}
\affiliation{Department of Physics, The Pennsylvania State University, University Park, Pennsylvania, 16802, USA}
\affiliation{Center for Theory of Emergent Quantum Matter, The Pennsylvania State University, University Park, Pennsylvania 16802, USA}
\affiliation{Institute for Computational and Data Sciences, The Pennsylvania State University, University Park, Pennsylvania 16802, USA}
\affiliation{Materials Research Institute, The Pennsylvania State University, University Park, Pennsylvania 16802, USA}
\author{Bryce Gadway}
\email{bgadway@psu.edu}
\affiliation{Department of Physics, The Pennsylvania State University, University Park, Pennsylvania, 16802, USA}

\maketitle

\tableofcontents

\section{Derivation of the effective master equation}
\subsection{Our model in the frequency domain}
Our model consists of three parts: an interacting Rydberg atom system with negligible intrinsic dissipation that is in proximity to two types of auxiliary atoms (sources and sinks) that have engineered dissipation on their Rydberg states. The dissipative auxiliary atoms act to stabilize the target states of the system. In the following, we use the subscripts ``S", ``A", and ``B" to denote the system atoms, source atoms, and sink atoms, respectively.
Without loss of generality, we express the system Hamiltonian in its eigenbasis as
\begin{equation}
\hat{H}_S = \sum_{\lambda} \lambda\, |\lambda\rangle\langle \lambda| .
\end{equation}
A raising operator can be written as
\begin{equation}
\hat{S}^{+} = \sum_{\lambda,\lambda'} s_{\lambda\lambda'} |\lambda\rangle\langle\lambda'|, 
\qquad
\hat{S}^- = (\hat{S}^+)^\dagger ,
\end{equation}
with $s_{\lambda\lambda'}=\langle \lambda|\hat{S}^\dagger|\lambda'\rangle$. 
In the following, we use this general form of the raising operator $\hat{S}^{+}$, instead of a specific type of raising operator such as $\hat{S}_j^{+}=\hat{S}_j^x+i\hat{S}_j^y$, to keep the derivation general and simple.

As shown in the main text (Fig.~1), to induce controllable dissipation on the Rydberg states, each source and sink auxiliary atom consists of four internal states
$\{|g\rangle, |i\rangle, |0\rangle, |1\rangle\}$.
The two Rydberg states $|0\rangle$ and $|1\rangle$ are resonantly coupled to the ground state $|g\rangle$ and low-lying state $|i\rangle$, enabling engineered decay processes.
For each source atom $j$, the Hamiltonian and jump operator are
\begin{equation}
\begin{aligned}
\hat{H}_{A,j}(t) &= 
\Delta_{A,j} |1\rangle_j\langle 1|
+ \left(\Omega_{i0,j}|i\rangle_j\langle0|
+ \Omega_{g1,j}e^{i\nu_{g1,j}t}|g\rangle_j\langle1|
+ \text{h.c.}\right), \\
\\
\hat{H}_{SA,j} &= J_j \big(|0\rangle_j\langle1|\otimes \hat{S}^+
+ |1\rangle_j\langle0|\otimes \hat{S}^-\big), \\
\\
\hat{L}_{A,j} &= \sqrt{\Gamma_j}\, |g\rangle_j\langle i| \otimes \hat{\mathbb I}_S .
\end{aligned}
\end{equation}
Here, the detunings are defined as relative frequency offsets with respect to the natural frequency.
Similarly, for each sink atom $k$, 
\begin{equation}
\begin{aligned}
\hat{H}_{B,k}(t) &= -\Delta_{B,k} |0\rangle_k\langle0|
+ \left(\Omega_{i1,k}|i\rangle_k\langle1| 
+ \Omega_{g0,k} e^{i\nu_{g0,k}t}|g\rangle_k\langle0|
+\text{h.c.}\right), \\
\\
\hat{H}_{SB,k} &= J_k \big(|0\rangle_k\langle1|\otimes \hat{S}^+
+ |1\rangle_k\langle0|\otimes \hat{S}^-\big), \\
\\
\hat{L}_{B,k} &= \sqrt{\Gamma_k}\, |g\rangle_k\langle i| \otimes \hat{\mathbb I}_S .
\end{aligned}
\end{equation}
The total Hamiltonian reads
\begin{equation}
\hat{H}(t) = \hat{H}_S + \sum_j \!\left(\hat{H}_{A,j}(t)+\hat{H}_{SA,j}\right)
+ \sum_k \!\left(\hat{H}_{B,k}(t)+\hat{H}_{SB,k}\right),
\end{equation}
and the dynamics is governed by the Lindblad master equation
\begin{equation}
\dot{\hat{\rho}} = -i[\hat{H}(t), \hat{\rho}]
+ \sum_j \hat{\mathcal D}[\hat{L}_{A,j}]\hat{\rho}
+ \sum_k \hat{\mathcal D}[\hat{L}_{B,k}]\hat{\rho} ,
\end{equation}
where $\hat{\mathcal D}[c]\hat{\rho}=c\hat{\rho} c^\dagger-\tfrac{1}{2}\{c^\dagger c,\hat{\rho}\}$.

To get a clear physical picture on the stabilization mechanism, we now perform a rotating-frame transformation and remove the unnecessary parts.
We define
\begin{equation}
\hat{H}_0 \equiv \hat{H}_S 
+ \sum_j \Delta_{A,j}|1\rangle_j\langle1|
- \sum_k \Delta_{B,k}|0\rangle_k\langle0| ,
\qquad
\hat{R}(t)=e^{i\hat{H}_0 t}.
\end{equation}
The Hamiltonian in the rotating frame ($H_{\rm rot}=\hat{R}\hat{H}\hat{R}^\dagger+i\dot{\hat{R}}\hat{R}^\dagger$) is
\begin{equation}
\hat{H}_{\rm rot}(t)=\hat{R}\big(\hat{H}(t)-\hat{H}_0\big)\hat{R}^\dagger
= \sum_j \Big(\hat{H}_{A,j}^{\rm rot}(t)+\hat{H}_{SA,j}^{\rm rot}(t)\Big)
+ \sum_k \Big(\hat{H}_{B,k}^{\rm rot}(t)+\hat{H}_{SB,k}^{\rm rot}(t)\Big).
\end{equation}
For each source atom $j$, the internal auxiliary Hamiltonian under resonant drive becomes
\begin{equation}\label{seq10}
\begin{aligned}
\hat{H}_{A,j}^{\rm rot}
= \Omega_{i0,j}\big(|i\rangle_j\langle0| + |0\rangle_j\langle i|\big) 
 + \Omega_{g1,j}\!\left(|g\rangle_j\langle1|
+ |1\rangle_j\langle g|\right),
\end{aligned}
\end{equation}
Using
$
\hat{R}|\lambda\rangle\langle\lambda'|\hat{R}^\dagger
= e^{i(\lambda-\lambda')t}|\lambda\rangle\langle\lambda'|
$,
the interaction Hamiltonian transforms to
\begin{equation}
\begin{aligned}
\hat{H}_{SA,j}^{\rm rot}(t)
&= J_j\Big(\hat{R}\,|0\rangle_j\langle1|\,\hat{R}^\dagger\Big)\otimes
\Big(\hat{R}\,\hat{S}^+\,\hat{R}^\dagger\Big) + \text{h.c.}\\
&= \sum_{\omega} J_j e^{i(\omega-\Delta_{A,j})t}
|0\rangle_j\langle1|\otimes \hat{S}^+(\omega) + \text{h.c.},
\end{aligned}
\end{equation}
where the spectral components are
\begin{equation}
\hat{S}^+(\omega)=\sum_{\lambda-\lambda'=\omega}
s_{j,\lambda\lambda'}|\lambda\rangle\langle\lambda'|,
\qquad
\hat{S}^-(\omega)=\left(\hat{S}^+(\omega)\right)^\dagger.
\end{equation}
Similarly, for each sink atom $k$, the internal auxiliary Hamiltonian becomes
\begin{equation}
\begin{aligned}
\hat{H}_{B,k}^{\rm rot}
= \Omega_{i1,k}\big(|i\rangle_k\langle1| + |1\rangle_k\langle i|\big) 
 + \Omega_{g0,k}\!\left(|g\rangle_k\langle0|
+ |0\rangle_k\langle g|\right).
\end{aligned}
\end{equation}
The interaction Hamiltonian transforms to
\begin{equation}
\begin{aligned}
\hat{H}_{SB,k}^{\rm rot}(t)
&= \sum_{\omega} J_k e^{-i(\omega-\Delta_{B,k})t}
|1\rangle_k\langle0|\otimes \hat{S}^-(\omega) + \text{h.c.}.
\end{aligned}
\end{equation}
Together, the master equation in this rotating frame reads
\begin{equation}
\begin{aligned}
    \dot{\hat{\rho}}
&= -i[\hat{H}_{\rm rot}(t),\hat{\rho}]
+ \sum_j \hat{\mathcal D}[\hat{L}_{A,j}]\hat{\rho}
+ \sum_k \hat{\mathcal D}[\hat{L}_{B,k}]\hat{\rho},
\\
\hat{H}_{\rm rot}(t)
&=
\sum_j \Big(\hat{H}_{A,j}^{\rm rot}+\hat{H}_{SA,j}^{\rm rot}(t)\Big)
+ \sum_k \Big(\hat{H}_{B,k}^{\rm rot}+\hat{H}_{SB,k}^{\rm rot}(t)\Big),
\\
\hat{H}_{A,j}^{\rm rot}
&= \Omega_{i0,j}\big(|i\rangle_j\langle0| + |0\rangle_j\langle i|\big) 
 + \Omega_{g1,j}\!\left(|g\rangle_j\langle1|
+ |1\rangle_j\langle g|\right),
\\
\hat{H}_{SA,j}^{\rm rot}(t)
&= \sum_{\omega} J_j e^{i(\omega-\Delta_{A,j})t}
|0\rangle_j\langle1|\otimes \hat{S}^+(\omega) + \text{h.c.},
\\
\hat{H}_{B,k}^{\rm rot}
&= \Omega_{i1,k}\big(|i\rangle_k\langle1| + |1\rangle_k\langle i|\big) 
 + \Omega_{g0,k}\!\left(|g\rangle_k\langle0|
+ |0\rangle_k\langle g|\right),
\\
\hat{H}_{SB,k}^{\rm rot}(t)
&= \sum_{\omega} J_k e^{-i(\omega-\Delta_{B,k})t}
|1\rangle_k\langle0|\otimes \hat{S}^-(\omega) + \text{h.c.}.
\end{aligned}
\end{equation}

\subsection{Effective Lindblad operators}
While the full master equation derived above is sufficient to stabilize the target states, it is also useful to derive a more concise form that makes the physical process more transparent. 
In this section, we give a reduced master equation using adiabatic elimination based on the effective operator formalism for open quantum systems~\cite{reiter2012}.

Under the adiabatic-elimination condition (large decay limit, assuming negligible population in the excited subspace $\mathcal{E}_j=\mathrm{span}\{|i\rangle_j,|0\rangle_j\}$ for the source auxiliaries), we apply the effective operator formalism~\cite{reiter2012}. The Hilbert space is decomposed into an excited subspace $\mathcal{E}_j$ and a ground subspace $\mathcal{G}_j=\mathrm{span}\{|g\rangle_j,|1\rangle_j\}$.
For the source atom (from Eq.~(\ref{seq10})), the excited subspace Hamiltonian is
\begin{equation}
\hat{H}^{\mathcal{E}}_{A,j} = \Omega_{i0,j}\big(|i\rangle_j\langle0| + |0\rangle_j\langle i|\big),
\end{equation}
and the ground subspace Hamiltonian is
\begin{equation}
\hat{H}^{\mathcal{G}}_{A,j} = \Omega_{g1,j}\big(|g\rangle_j\langle1| + |1\rangle_j\langle g|\big).
\end{equation}
The effective non-Hermitian Hamiltonian then takes the form of
\begin{equation}
\hat{H}^{\mathrm{NH}}_{A,j} = \hat{H}^{\mathcal{E}}_{A,j} - \tfrac{i}{2}\hat{L}_{A,j}^\dagger \hat{L}_{A,j}
=\Omega_{i0,j}\big(|i\rangle_j\langle0| + |0\rangle_j\langle i|\big) - \tfrac{i\Gamma_j}{2}|i\rangle_j\langle i|.
\end{equation}
Its inverse in the $\{|i\rangle_j,|0\rangle_j\}$ basis is
\begin{equation}
(\hat{H}^{\mathrm{NH}}_{A,j})^{-1}
= \begin{pmatrix}
0 & \Omega_{i0,j}^{-1} \\
\Omega_{i0,j}^{-1} & i\Gamma_j/(2\Omega_{i0,j}^2)
\end{pmatrix}.
\end{equation}
The excitation operator from $\mathcal{G}_j$ to $\mathcal{E}_j$ is
\begin{equation}
\hat V_{+,j}=\sum_{\omega} J_j\,e^{i(\omega-\Delta_{A,j})t}\,
\big(|0\rangle_j\langle1|\big)\otimes \hat S^{+}(\omega),
\qquad
\hat V_{-,j}=\hat V_{+,j}^\dagger .
\end{equation}
The effective operators acting on $\mathcal{G}_j\otimes \mathcal H_S$ are then~\cite{reiter2012}
\begin{equation}
\hat H_{A,j}^{\mathrm{eff}}
=
\hat H^{\mathcal{G}}_{A,j}
-\tfrac{1}{2}\hat V_{-,j}\Big[(\hat H^{\mathrm{NH}}_{A,j})^{-1}+((\hat H^{\mathrm{NH}}_{A,j})^{-1})^\dagger\Big]\hat V_{+,j},
\end{equation}
and
\begin{equation}
\hat L_{A,j}^{\mathrm{eff}}=\hat L_{A,j}\,(\hat H^{\mathrm{NH}}_{A,j})^{-1}\hat V_{+,j}.
\end{equation}
In the present level structure, the Hamiltonian correction term vanishes (since $(\hat H_{\mathrm{NH}}^{-1}+{\rm h.c.})$ has zero $\langle 0|\cdot|0\rangle$ element), such that
\begin{equation}
\hat H_{A,j}^{\mathrm{eff}}=\Omega_{g1,j}\big(|g\rangle_j\langle1| + |1\rangle_j\langle g|\big).
\end{equation}
Moreover, using $\hat L_{A,j}=\sqrt{\Gamma_j}\,|g\rangle_j\langle i|\otimes \hat{\mathbb I}_S$ and the above $(\hat H^{\mathrm{NH}}_{A,j})^{-1}$, we obtain
\begin{equation}
\hat L_{A,j}^{\mathrm{eff}}
=\sum_{\omega}\frac{J_j\sqrt{\Gamma_j}}{\Omega_{i0,j}}\,
e^{i(\omega-\Delta_{A,j})t}\,
\big(|g\rangle_j\langle1|\big)\otimes \hat S_j^{+}(\omega).
\end{equation}
For each sink atom $k$, we eliminate $\{|1\rangle_k,|i\rangle_k\}$ while keeping $\{|g\rangle_k,|0\rangle_k\}$.
The excited subspace $\mathcal E_k=\mathrm{span}\{|1\rangle_k,|i\rangle_k\}$ is governed by
\begin{equation}
\hat H^{\mathrm{NH}}_{B,k}=
\Omega_{i1,k}\big(|i\rangle_k\langle1| + |1\rangle_k\langle i|\big)
-\tfrac{i\Gamma_k}{2}|i\rangle_k\langle i|.
\end{equation}
Its inverse in the $\{|1\rangle_k,|i\rangle_k\}$ basis is
\begin{equation}
(\hat H^{\mathrm{NH}}_{B,k})^{-1}
=
\begin{pmatrix}
i\Gamma_k/(2\Omega_{i1,k}^2) & \Omega_{i1,k}^{-1} \\
\Omega_{i1,k}^{-1} & 0
\end{pmatrix}.
\end{equation}
The effective Hamiltonian on $\{|g\rangle_k,|0\rangle_k\}\otimes\mathcal H_S$ is
\begin{equation}
\hat H_{B,k}^{\mathrm{eff}}=
\Omega_{g0,k}\big(|g\rangle_k\langle0| + |0\rangle_k\langle g|\big),
\end{equation}
and the effective jump operator becomes
\begin{equation}
\hat L_{B,k}^{\mathrm{eff}}
=\sum_{\omega}\frac{J_k\sqrt{\Gamma_k}}{\Omega_{i1,k}}\,
e^{-i(\omega-\Delta_{B,k})t}\,
\big(|g\rangle_k\langle0|\big)\otimes \hat{S}^{-}(\omega).
\end{equation}
Then the effective master equation reads
\begin{equation}\label{eq23-eff}
\begin{aligned}
\dot{\hat\rho} &= -i [\hat H_{\mathrm{eff}}, \hat\rho]
+ \sum_j \hat{\mathcal D}[\hat L_{A,j}^{\mathrm{eff}}]\hat\rho
+ \sum_k \hat{\mathcal D}[\hat L_{B,k}^{\mathrm{eff}}]\hat\rho, \\
\hat H_{\mathrm{eff}} &=
\sum_j \Omega_{g1,j}\big(|g\rangle_j\langle1| + |1\rangle_j\langle g|\big)\otimes \hat{\mathbb I}_S
+ \sum_k \Omega_{g0,k}\big(|g\rangle_k\langle0| + |0\rangle_k\langle g|\big)\otimes \hat{\mathbb I}_S, \\
\hat L_{A,j}^{\mathrm{eff}} &= \sum_{\omega}A(\omega)\,
\big(|g\rangle_j\langle1|\big)\otimes \hat{S}^{+}(\omega), \\
\hat L_{B,k}^{\mathrm{eff}} &= \sum_{\omega}
B(\omega)
\big(|g\rangle_k\langle0|\big)\otimes \hat{S}^{-}(\omega).
\end{aligned}
\end{equation}
where 
$A(\omega)=\frac{J_j\sqrt{\Gamma_j}}{\Omega_{i0,j}}\,
e^{i(\omega-\Delta_{A,j})t}$ and 
$B(\omega)=\frac{J_k\sqrt{\Gamma_k}}{\Omega_{i1,k}}\,
e^{-i(\omega-\Delta_{B,k})t}$ imply that the coupling is resonantly enhanced near
$\omega\simeq \Delta_{A,j}$ or $\Delta_{B,k}$.

\begin{figure}[tb]
    \centering
    \includegraphics[width=1\linewidth]{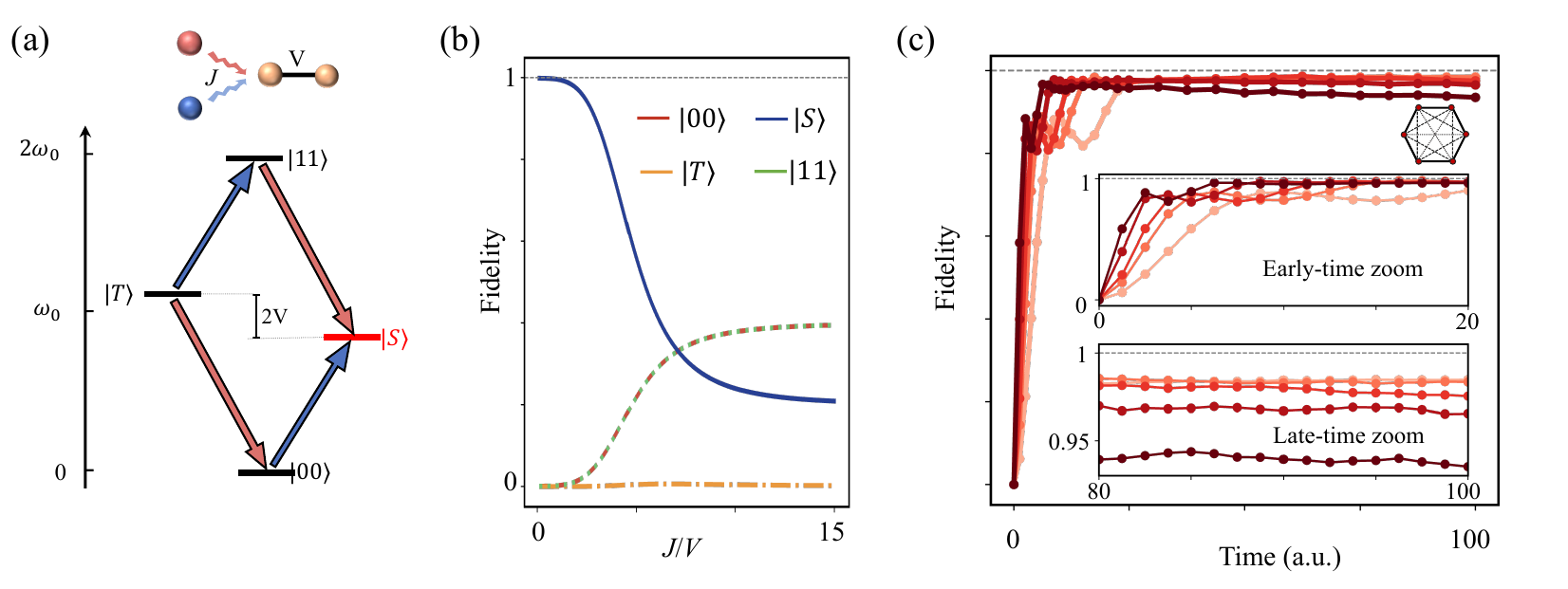}
    \caption{Nonreciprocal single-particle excitation and de-excitation.
    (a)~Stabilization of an entangled singlet state in a two-atom system via energy-selective (de-)excitations.
    The arrow shows the nonreciprocal transition path for stabilization of the singlet state, where the red path is realized via resonant coupling to a ``source'' atom, while the blue path arises from coupling to a ``sink'' atom. 
    The fidelity between the steady state and four eigenstates as a function of the system--auxiliary coupling strength are shown in (b). Here, both auxiliary atoms are chosen with the same parameters ($\gamma$, $\Omega$, $J$), and the effective decay rate is fixed at $\gamma/V = 0.05$, $\Omega/V = 0.05$.
    (c)~Time evolution of the fidelity between the prepared state and the $n=1$ ground state of the dipolar hexagonal $XY$ model. Darker colors correspond to larger values of $J$, with $J=[0.5,0.8,1,1.5,2]$. Smaller $J$ leads to higher fidelity but requires a longer preparation time, while larger $J$ accelerates the dynamics at the cost of reduced fidelity.
    The other parameters are $\gamma/J=0.65$, $\Omega/J=0.12$, and $V=20$.
    }
    \label{sfig1}
\end{figure}

\section{Protocol Performance}
Equation~(\ref{eq23-eff}) provides a simplified effective Lindblad operator that captures the mechanism of nonreciprocal pumping and depumping processes. 
It also indicates the characteristic timescale for a single excitation process at frequency $\omega$: 
\[
\tau \sim 1/|A(\omega)|^2,
\]
If an $n$-particle target state is prepared sequentially from the vacuum through $n$ single-particle steps, the total preparation time scales as $O(n\tau)$. 
This timescale can be reduced, e.g., by choosing an initial state closer to the target sector or by employing many embedded source/sink auxiliaries. 
For example, in a quadratic fermionic-like Hamiltonian where each quasiparticle can be independently excited by a dedicated source auxiliary, the preparation time for the $n$-filled state remains on the order of $O(\tau)$.

We note that the above analysis assumes the adiabatic limit, which is not essential in practice. Away from this limit, an analytical treatment is no longer available. Due to the computational complexity of simulating the full superoperator dynamics, our numerical study is restricted to small system sizes and a minimal number of auxiliary atoms (one, two, or three). Nevertheless, we can formulate general guiding principles for the performance of the protocol beyond the adiabatic regime.
Qualitatively, the dissipative preparation can be viewed as a directed walk within the system Hilbert space. Ultimately, the transition rates within this Hilbert space are limited by the auxiliary-system coupling rate $J$. To approach the desired target state quickly, one should want to work with a large initial $J$. 
At the same time, high-fidelity stabilization requires the auxiliary parameters, such as $J$ and $\gamma$, to remain perturbative relative to the intrinsic many-body energy scale (typically $J,\gamma \ll V$), so that the steady state remains close to the target eigenstate of the bare system Hamiltonian. 
Therefore, in practice there exists a tradeoff between preparation speed and final-state fidelity.

This tradeoff can be illustrated using a simple two-atom system, where the steady state can be obtained exactly. 
As shown in Fig.~\ref{sfig1}(a), the source and sink auxiliaries are tuned to complementary transition frequencies to stabilize the singlet state. 
Figure~\ref{sfig1}(b) shows that the fidelity with the target singlet decreases once the coupling ratio $J/V$ becomes too large, consistent with the expectation that strong auxiliary coupling perturbs the bare-system eigenstructure.

Extending this analysis to a hexagon XY model with dipolar interactions  (isotropic in sign, but depending on interparticle distance $r$ as $1/r^3$), we numerically study the relation between fidelity and preparation time for different values of $J/V$. 
As shown in Fig.~\ref{sfig1}(c), smaller $J/V$ leads to higher fidelity but requires a longer preparation time, while larger $J/V$ accelerates the dynamics at the cost of reduced fidelity. 
To balance these competing requirements, we perform a Bayesian optimization~\cite{bayesian} over the parameter space $J/V, \Omega/V, \gamma/V \in (0,0.1]$. The parameters used in the main text are selected from this optimization, based on a total of $100$ iterations, yielding a near-optimal tradeoff between rapid state preparation and high steady-state fidelity.

Alternatively, one may employ multiple auxiliaries with different fixed detunings, which could enable parallel excitation and relaxation processes and thereby further accelerate the preparation dynamics and reduce the overall preparation time. 
The time to reach the target state is also influenced by the choice of initial state (i.e., its projection onto system eigenstates and their distance in Hilbert space). Thus, one may consider sampling different initial product states of the system, based on a combination of global and local single-particle control, to expedite the convergence to the final state.

\section{Examples of stabilizing quantum states}
\subsection{Ground states in Bose-Hubbard model}
\begin{figure}[tb]
    \centering
    \includegraphics[width=1\linewidth]{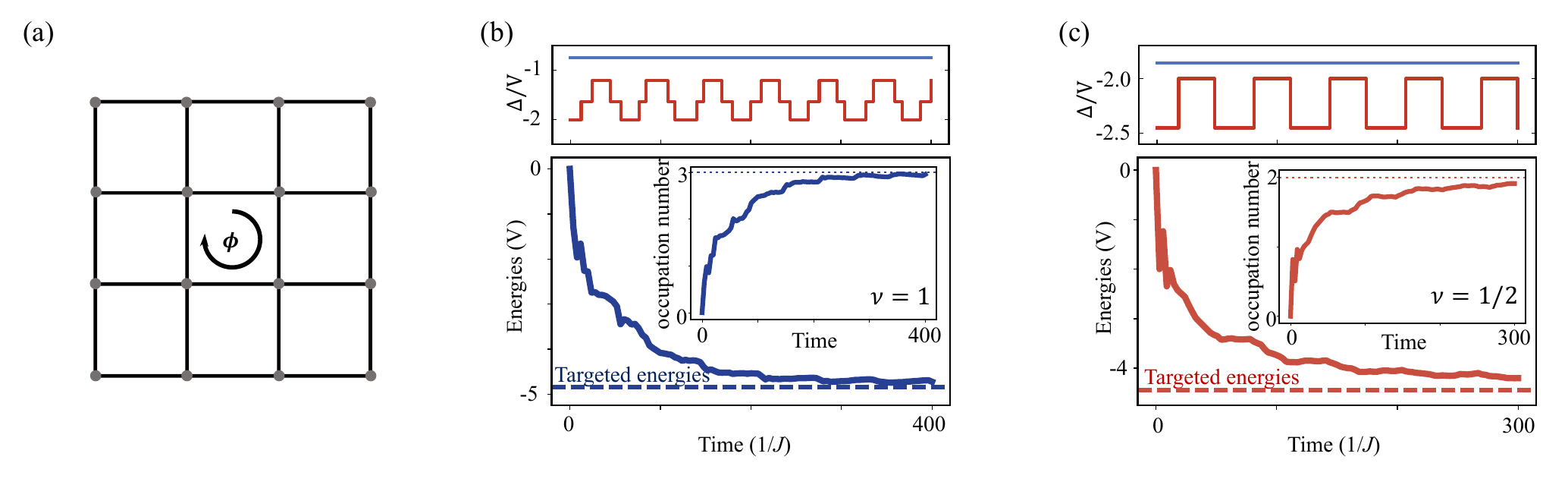}
    \caption{Ground-state preparation in a hard-core Bose-Hubbard model with effective magnetic flux. 
    (a)~A two-dimensional Bose--Hubbard model in the strong-interaction limit, which reduces to the Harper--Hofstadter model of hard-core bosons on a square lattice with magnetic flux $\alpha$. Such a model can be realized using dipolar XY interactions, with complex hopping phases induced by Floquet control. 
    (b)~Detuning protocol and state preparation for a $3\times 3$ torus with $\alpha=1/3$. The lower panel shows the time evolution of the prepared-state energy, which approaches the targeted ground-state energy during the dynamics. The inset shows the corresponding particle number, indicating a filling $\nu=1$, corresponding to an integer quantum Hall-like state. 
    (c)~Detuning protocol and state preparation for a $4\times 3$ torus with $\alpha=1/3$. The lower panel shows the time evolution of the prepared-state energy, and the inset gives the corresponding particle number. The target state has filling $\nu=1/2$, corresponding to a fractional quantum Hall-like (FQH) state, although for this small system size, it should not be viewed as a true FQH state. 
    In both cases, the prepared-state energy approaches the targeted state during the evolution. Parameters: $\gamma/J=0.3$, $\Omega/J=0.2$, $V/J=20$.
The dynamics are simulated using the quantum trajectory method with $1000$ trajectories.}
    \label{sfig2}
\end{figure}

In the main text, we numerically verified the efficiency of the stabilization protocol for preparing ground states in the dipolar spin-1/2 XY model. 
As a complementary example, here we show that the same idea can also be applied to a strongly interacting bosonic system with magnetic flux. 
Specifically, we consider a two-dimensional Bose-Hubbard model on a square lattice, described by
\begin{equation}\label{seq30}
\hat{H}_{\rm BH}
=
V\sum_{x,y}
\left(
e^{i2\pi\alpha y}\hat{a}_{x+1,y}^{\dagger}\hat{a}_{x,y}
+
\hat{a}_{x,y+1}^{\dagger}\hat{a}_{x,y}
+\mathrm{h.c.}
\right)
+\frac{U}{2}\sum_j \hat{n}_j(\hat{n}_j-1),
\end{equation}
where $\hat{a}_{x,y}$ is the boson annihilation operator at site $(x,y)$ and $\hat{n}_j=\hat{a}_j^\dagger \hat{a}_j$ is the number operator. 
Here $\alpha$ denotes the magnetic flux per plaquette through the Peierls phase in the hopping term.

In the strong-interaction regime $U\gg V$, the system is driven into the hard-core boson limit, where double occupancy is suppressed. 
The model then reduces to the Harper--Hofstadter Hamiltonian of hard-core bosons,
\begin{equation}
\hat{H}_{\rm Harper}
=
V\sum_{x,y}
\left(
e^{i2\pi\alpha y}\hat{a}_{x+1,y}^{\dagger}\hat{a}_{x,y}
+
\hat{a}_{x,y+1}^{\dagger}\hat{a}_{x,y}
+\mathrm{h.c.}
\right),
\end{equation}
which can be realized using dipolar XY interactions with Floquet-engineered complex hopping phases~\cite{yang2022}.

Despite the much more complex many-body spectrum of this interacting system, the same frequency-selective stabilization principle remains applicable, without requiring detailed \textit{a priori} spectral knowledge.
Due to the limitation of exact numerics, we consider small torus geometries. 
As shown in Fig.~\ref{sfig2}(b), we study a $3\times 3$ torus at flux $\alpha=1/3$, where the target state has filling $\nu=1$. 
The corresponding ground state is an integer quantum Hall-like state. 
Figure~\ref{sfig2}(c) shows a $3\times 4$ torus, also at $\alpha=1/3$, with filling $\nu=1/2$. 
In this case, the ground state corresponds to a fractional quantum Hall-like state, although for such a small system size it should be regarded only as a finite-size precursor rather than a true FQH state.

In both cases, the upper panels show the detuning protocols used for the auxiliary atoms, while the lower panels show the time evolution of the prepared-state energy. 
The prepared-state energy approaches the exact ground-state energy during the evolution, therefore driving the system into the target many-body state. 
The insets further show the evolution of the particle number, which approaches the desired filling in each case.

\subsection{Multiple steady states in the Ising model}
\begin{figure}
    \centering
    \includegraphics[width=1\linewidth]{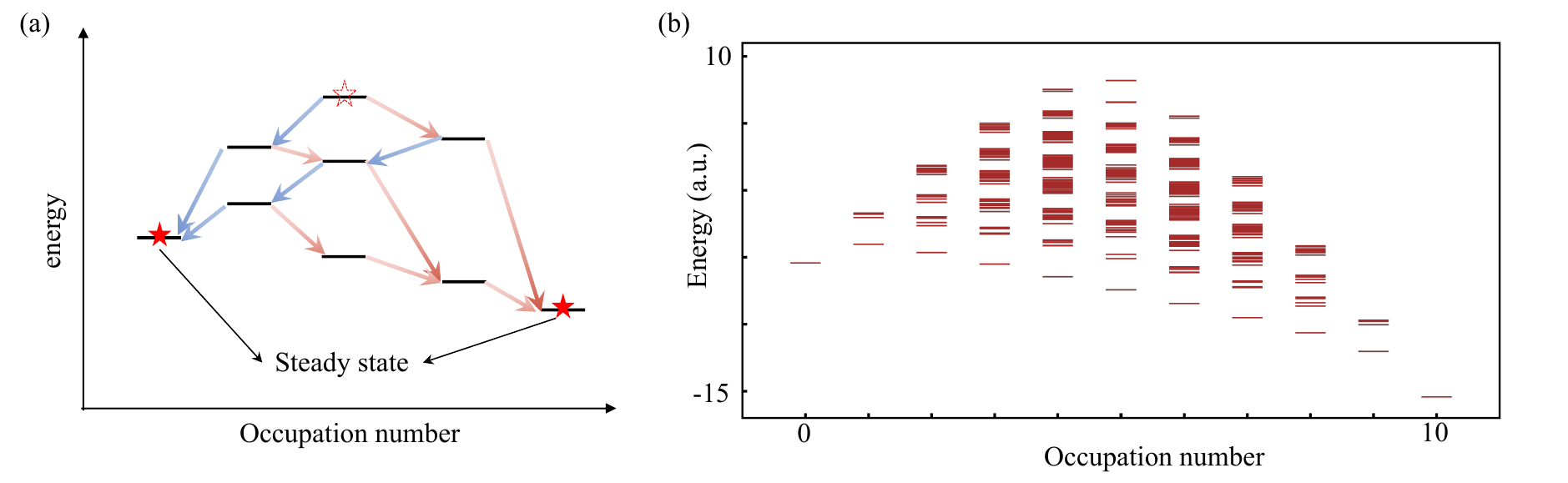}
\caption{Multiple steady states in frequency-selective dissipative dynamics. (a)~Schematic illustration of bistability in energy--particle-number space. Red (blue) arrows denote excitation (de-excitation) processes induced by source (sink) auxiliaries. For a non-convex energy landscape, multiple local minima can act as stabilizers. (b)~Energy spectrum of a long-range Ising model in a longitudinal field for a one-dimensional chain. The non-monotonic dependence of energy on particle number gives rise to local minima on two sides. The spectra (natural frequency frame) are obtained by using $N=10$, $V=-1$, $\Delta=0.5$, and $r_{ij}=1$ for nearest neighbor.}
    \label{sfig4}
\end{figure}

The examples discussed in the main text and above focus on cases where the engineered dissipative dynamics converges to a steady ground state at a unique filling. 
This occurs when the frequency-selective excitation and de-excitation processes guide the system toward a unique target eigenstate. Here, we discuss cases of multiple steady states residing in distinct filling sectors.

Generally, the steady states depend on the interplay between the system spectrum and the frequency selectivity of the auxiliary-induced transitions. When the energy as a function of particle number is non-convex---i.e., it contains multiple local minima---and transitions may in practice be restricted to these ``basins,'' as sketched in Fig.~\ref{sfig4}(a).

A representative example is the Ising model in a longitudinal field,
\begin{equation}
\hat{H}_{\rm Ising}=\sum_{ij}\frac{V}{r_{ij}^3}S_i^z S_j^z+\Delta \sum_i S_i^z,
\label{seq_ising}
\end{equation}
whose spectrum, shown in Fig.~\ref{sfig4}(b), exhibits a non-monotonic dependence on the magnetization. In this case, the initial states can evolve on intermediate timescales toward different eigenstates with local minima energies.
This behavior extends the protocol beyond single-state preparation: by tuning the auxiliary spectral response and choosing appropriate initial conditions, one may selectively stabilize different many-body states. Such sampling-based explorations will be important for studying the metastability inherent to many spin models (e.g., the false vacuum of Eq.~(\ref{seq_ising}) and the metastable landscape of spin glass models).

\section{Numerical method and error analysis}
\begin{figure}[tb]
    \centering
    \includegraphics[width=0.9\linewidth]{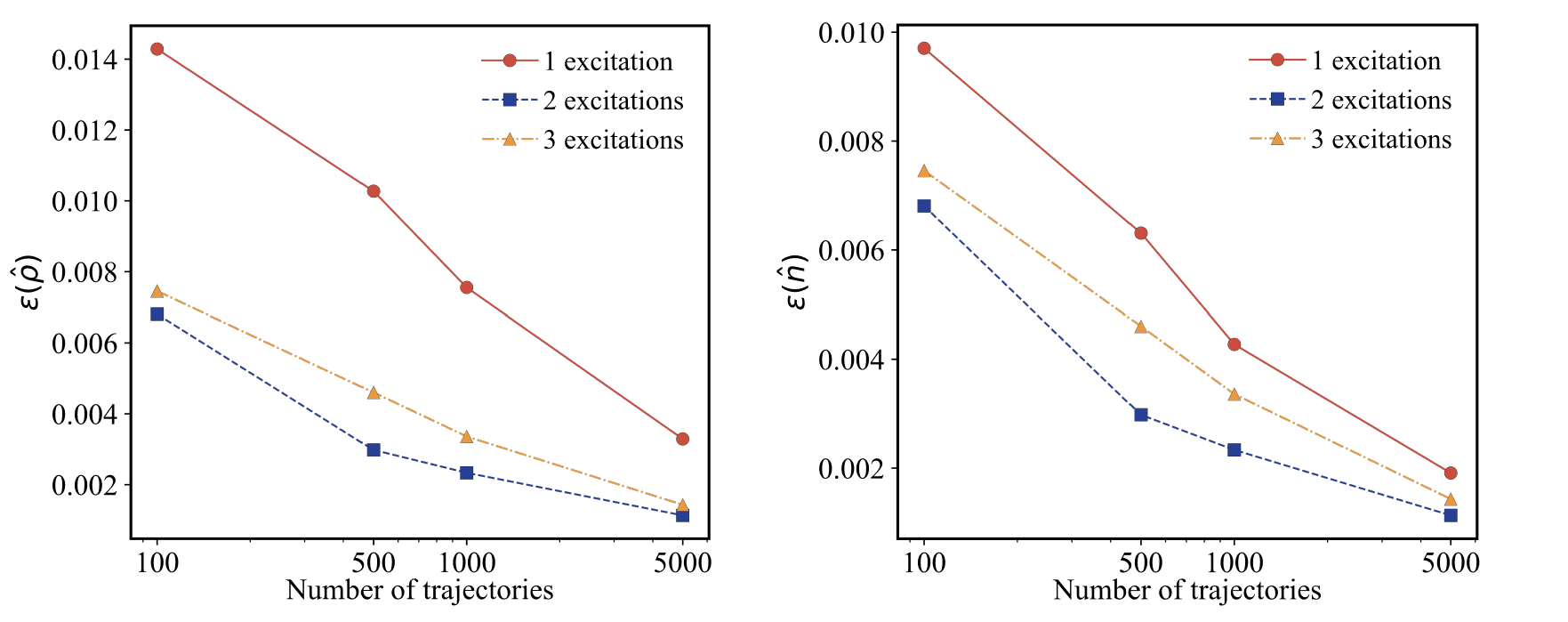}
    \caption{Error analysis of the uncertainty in quantum trajectories. 
    The total density matrix (a) and occupation number (b) are used as the observables to evaluate the errors. The data we used is given by Fig.~2(d) in the main text.}
    \label{sfig3}
\end{figure}

In our system, each sink and source atom has four internal states, while each system atom is considered to have two states (thus ignoring the intrinsic spontaneous emission of the excited Rydberg states).
The total Hilbert space dimension is therefore
$
N_H = 4^M 2^N,
$
and the corresponding density matrix lives in a space of dimension
\begin{equation}
    N_H^2 = (4^M 2^N)^2.
\end{equation}
This scaling quickly becomes numerically challenging even for relatively small system sizes (e.g., $N<16$). 
To reduce the computational cost, we use the quantum trajectory (Monte Carlo wave function) method~\cite{daley2014}. 
In this approach, the open-system dynamics is unraveled into stochastic realizations of pure-state evolutions. 
Instead of propagating the full density matrix in time, which requires storing and evolving an object of size $N_H^2$, 
we propagate state vectors of size $N_H$ and perform a stochastic average over individual trajectories.

The results obtained from quantum trajectories are statistical and carry sampling uncertainty. 
For an observable $A$, the expectation value $\langle A \rangle$ is estimated from $N$ trajectories. 
The statistical error of this estimate is given by
\[
\sigma_A = \frac{\Delta_A}{\sqrt{N}},
\]
where $\Delta_A$ is the sample standard deviation of $A$ over the trajectories. 
The number of trajectories required for convergence depends on both the details of the dynamics and the observable being evaluated. 
For observables with non-zero mean, we typically require
\begin{equation}
   \epsilon (A)=\frac{\sigma_A}{\langle A \rangle} 
= \frac{\Delta_A}{\sqrt{N}\,\langle A \rangle} \ll 1,
\end{equation}
which sets a practical criterion for convergence (see our error data in Fig.~\ref{sfig3}).


%